# Spatial Nonlocality from Finite Extinction Ratio in BOTDR over Heterogeneous Fiber Links

Giannis Poulopoulos*, Constantinos Haliotis, Panagiotis Toumasis, Hercules Avramopoulos

***Abstract*— Finite extinction ratio in pulse-modulated Brillouin optical time-domain reflectometry (BOTDR) produces residual inter-pulse Brillouin backscatter outside the nominal pulse gate. We model and experimentally quantify this effect as spatial nonlocality in a heterogeneous fiber link and define the spectral contamination ratio (SCR) as the contaminating Brillouin-band power normalized to the local Brillouin-band power. Measurements were performed using a 1.2 km standard single-mode fiber (SMF) followed by 0.1 km of dispersion-compensating fiber (DCF), whose separated Brillouin frequency shifts allow section-to-section leakage to be resolved spectrally. At 20 m spatial resolution, increasing the extinction ratio from 18.5 to 33.0 dB reduced the SMF-band contamination in the DCF gates from −5.3 to −14.8 dB. Over the 16–33 dB range, the SCR showed fitted slopes of −1.064 and −1.246 dB/dB for the two contamination directions. These results quantify finite-ER-induced spatial nonlocality and provide a link-specific metric for extinction-ratio selection in BOTDR.**

## I. Introduction

Single-ended BOTDR measures distributed temperature or strain through spontaneous Brillouin backscatter generated by a pulsed probe [1]. Its one-end access is relevant to installed and multifunctional sensing links [2],[3]. Spatial localization relies on short probe pulses, which reduce the energy per resolution cell and increase sensitivity to receiver noise and pulse extinction [4]. With finite extinction ratio, residual inter-pulse power generates Brillouin backscatter outside the nominal pulse gate, introducing a spatial-locality error [5].

Finite extinction ratio has been studied mainly as a BOTDR performance limitation. A prior BOTDR analysis modeled the non-ideal modulated probe as the sum of a useful pulsed component and a residual leakage component, showing that the leakage-induced Brillouin signal degrades the SNR even after subtraction of the leakage floor [5], and demonstrated increase of the maximum temperature uncertainty from 1.9 °C to 5.6 °C by reducing the extinction ratio from 50 dB to 25 dB [5]. A subsequent high-extinction-ratio BOTDR implementation increased the probe extinction ratio from 35 dB to 65 dB using a synchronous optical-switch-based modulator, reducing the maximum temperature uncertainty from 5.2 °C to 0.8 °C [6]. For heterogeneous sensing links, the same experiment showed that leakage accumulated along a longer fiber section can contribute to the baseline measured in a shorter section with a different Brillouin frequency shift, thereby biasing the extracted response [6]. These studies did not, however, quantify this leakage as a directional section-to-section contribution.

Building on previous finite-ER studies, we examine spatial locality and not global SNR or temperature uncertainty. A heterogeneous SMF/DCF link with distinct Brillouin frequency shifts allows section-to-section leakage to be resolved spectrally. We define the spectral contamination ratio (SCR) as the contaminating Brillouin-band power normalized to the local band power within the same spatial gate and use its ER dependence to quantify finite-ER-induced spatial nonlocality in BOTDR.

## II. Concept

Short-pulse BOTDR improves spatial resolution but reduces the pulse energy available in each resolution cell [4]. With a finite-extinction modulator, the probe consists of a pulsed on-state and a residual inter-pulse component, so the launched power can be written as [5]:

$$P(t) = \begin{cases} P_{on}, & 0 < t < \tau \\ P_{off}, & \tau < t < T \end{cases} \quad (1)$$

where $P_{on}$ is the optical power during the pulse, $P_{off}$ is the residual inter-pulse power, and $T$ is the pulse period. The Extinction Ratio ($ER$) is defined by $ER = P_{\text{on}}/P_{\text{off}}$, and the Duty Cycle by $D = \tau/T$ .Over one period, the launched pulse energy is $E_{\text{on}} = P_{\text{on}}\tau$ , whereas the residual inter-pulse energy is $E_{\text{off}} = P_{\text{off}}(T-\tau)$. Their ratio is therefore:

$$\frac{E_{off}}{E_{on}} = \frac{T-\tau}{\tau} \cdot \frac{1}{ER} \quad (2)$$

Thus, for a fixed extinction ratio, the relative leakage energy increases as the duty cycle is reduced, which links pulse operation to leakage-induced nonlocal Brillouin scatter.

For a spatial gate assigned to section $j$, the measured Brillouin power can be written as the sum of a local pulse-induced term and a leakage-induced term generated by the residual inter-pulse field [5]. The local contribution is produced within the pulse-defined resolution cell, $\Delta z = c\tau/(2n)$, and scales as $P_{\text{local},j} \propto P_{\text{on}} G_j(f_j)\Delta z$, where $G_j(f_j)$ is the effective integrated Brillouin response of section $j$, and is approximated by $G_j(f_j) \simeq \rho_j \kappa_j A_j g_{B,j}(f_j)\Delta f_j L_j^{eff}$, where $g_{B,m}$ is the Brillouin response, $\Delta f_j$ is the RF integration bandwidth, $L_j^{eff}$ is the effective interaction length, $A_j$ is the round-trip attenuation factor, $\rho_j$ accounts for polarization-dependent reception, and $\kappa_j$ for finite pulse-edge and residual waveform weighting. The leakage contribution at a contaminating frequency $f_{XT}$ is generated over the effective lengths of the fiber sections that contribute at that frequency and scales as $P_{\text{leak}}(f_{XT}) \propto P_{\text{off}} \sum_{i\neq j} G_i(f_{XT}) L_i^{\text{eff}}$.

The spectral contamination ratio (SCR) is defined as the leakage-induced Brillouin power at the contaminating

*Correspondence: Giannis Poulopoulos (jpoul@mail.ntua.gr).*
All authors are with Photonics Communications Research Laboratory, National Technical University of Athens, Patission 42., Athens, Greece.

frequency $f_{XT}$, normalized to the local Brillouin power measured in the same spatial gate at the local resonance frequency $f_j$:

$$SCR_j(f_{XT}) = \frac{P_{\text{leak}}(f_{XT})}{P_{\text{local},j}(f_j)} \propto \frac{1}{ER} \frac{\sum_{i\neq j} G_i\,(f_{XT}) L_i^{\text{eff}}}{G_j(f_j)\Delta z} \qquad (3)$$

In logarithmic form:

$$SCR_{j,\text{dB}}(f_{XT}) \simeq -ER_{\text{dB}} + 10\log_{10}\left[\frac{\sum_{i\neq j} G_i\,(f_{XT}) L_i^{\text{eff}}}{G_j(f_j)\Delta z}\right] \qquad (4)$$

For a fixed pulse duration and link composition, the model predicts an approximately linear $SCR$ dependence on extinction ratio with a slope of $-1dB/dB$. The second term gives the offset imposed by the effective lengths and relative Brillouin responses of the fiber sections.

In the two-section link, the measured terms are the SMF-band leakage in DCF gates and the DCF-band leakage in SMF gates. Their separated BFS values allow both contributions to be extracted by band integration. Same-BFS leakage remains spectrally overlapped with the local response and is not resolved by direct SCR extraction. For the two measured contamination terms, the model gives:

$$\begin{cases} SCR_{\text{SMF}\to\text{DCF}} \propto ER^{-1} G_{\text{SMF}} L_{\text{SMF}}^{\text{eff}} / (G_{\text{DCF}}\Delta z) \\ SCR_{\text{DCF}\to\text{SMF}} \propto ER^{-1} G_{\text{DCF}} L_{\text{DCF}}^{\text{eff}} / (G_{\text{SMF}}\Delta z) \end{cases} \qquad (5)$$

The $ER^{-1}$ factor gives the expected $-1\text{dB/dB}$ dependence when the link and acquisition conditions remain fixed. The effective section lengths and integrated Brillouin responses, including the fiber-dependent parameters defined above, determine the absolute SCR levels and the difference between the two contamination values, and not the $ER^{-1}$ factor.

## III. Experimental setup and results

The experimental setup used to verify the section-to-section contamination model is shown in Figure 1. A 1550 nm narrow-linewidth laser is split into probe and LO branches. Two cascaded MZMs generate 200 ns pulses at 50 kHz and provide enhanced inter-pulse suppression. The probe is amplified to 16 dBm, filtered by a 1 nm OBPF, and launched into 1.2 km of SMF followed by 0.1 km of DCF. The sections have BFS values of 10.85 and 9.62 GHz and linewidths of 30 and 45 MHz, respectively. The backscatter is detected by a 30 GHz photoreceiver and recorded at 40 GSa/s. The 200 ns pulse gives approximately 20 m spatial resolution.

The digitized time traces were captured from the oscilloscope, aligned to the pulse trigger, and parsed into spatial gates using a block size of 8000 samples. The spatial resolution is determined by the pulse duration according to $\Delta z = c\tau/(2n)$, resulting in $\Delta z \approx 20$m for $\tau = 200$ns, and thus 100 spatial gates are defined per pulse period. A Hann window is applied to each block to suppress spectral leakage. Since spontaneous Brillouin scattering has random phase between acquisitions, the spectra were averaged in power, and the power spectral density (PSD) was obtained by averaging $|FFT|^2$ over $N = 25$ acquisitions for each gate.

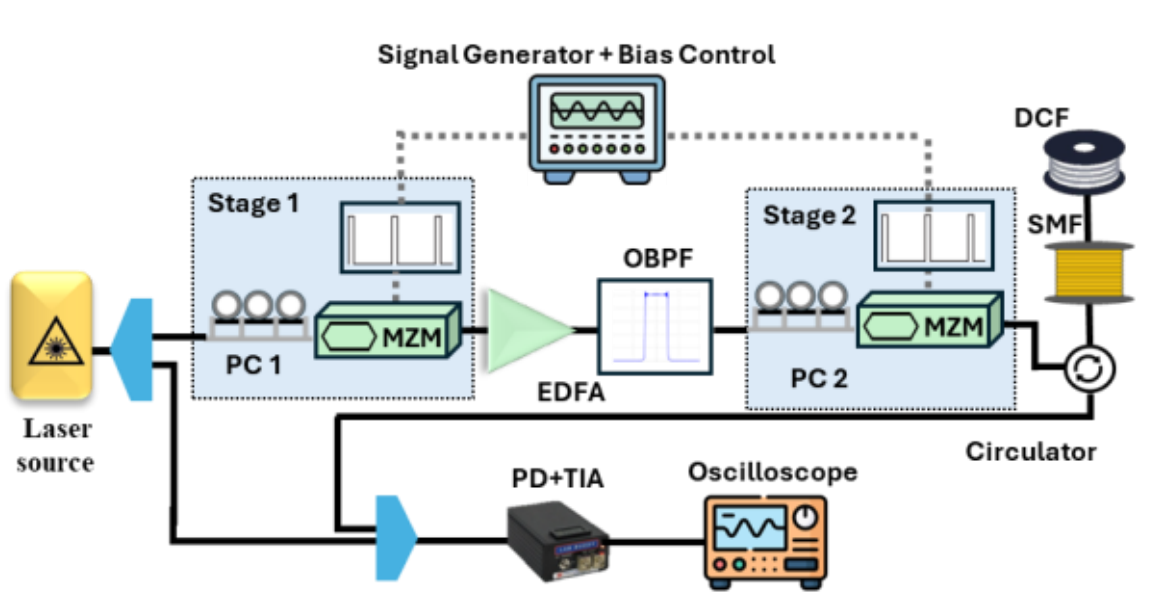

Figure 1. Experimental BOTDR testbed

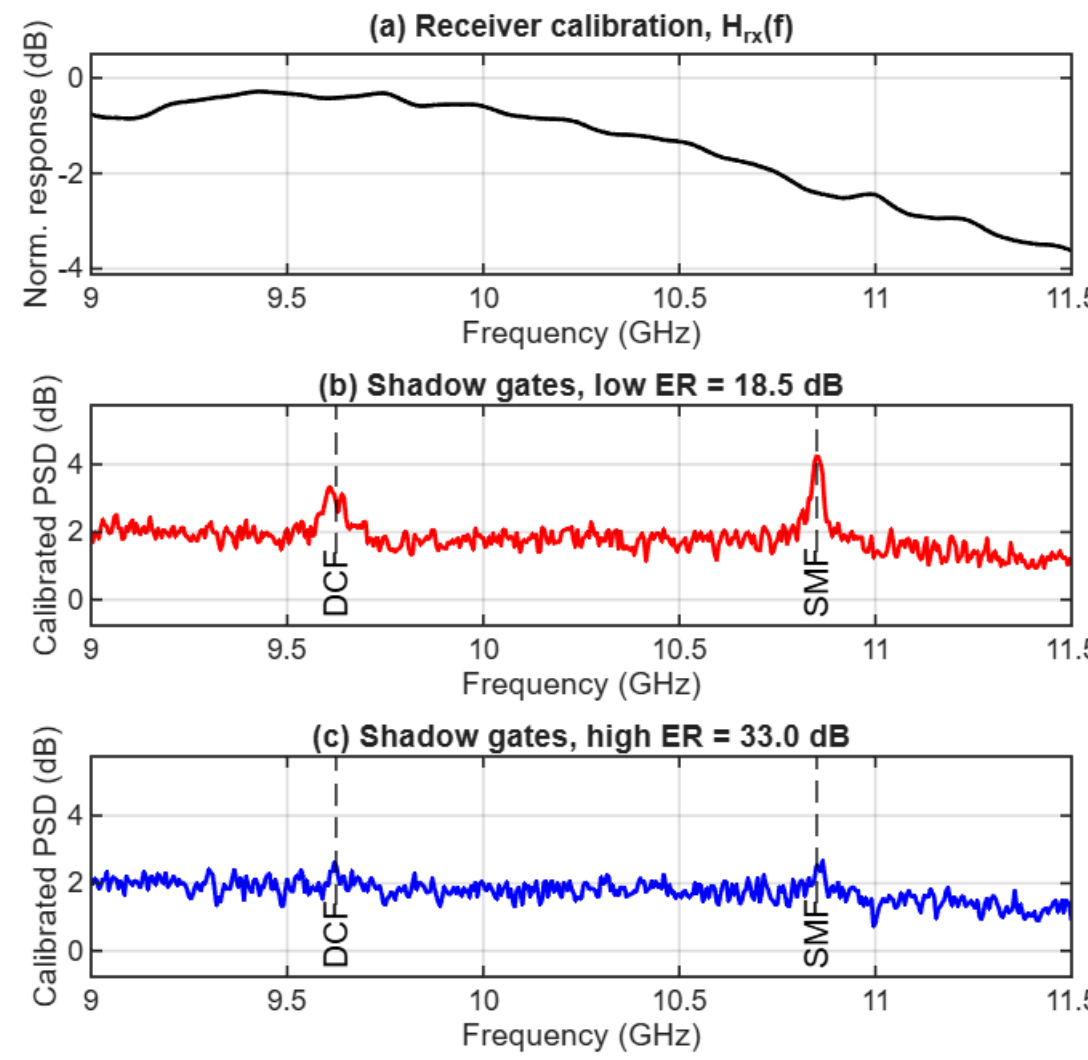

Figure 2. (a) $H_{\text{rx}}(f)$ Receiver calibration, Shadow-gate spectrum under (b) low-ER and (c) high-ER operation.

The signal processing begins by characterizing the frequency response of the receiver hardware, $H_{\text{rx}}$ (Figure 1.a), using a local-oscillator-only acquisition with the probe path and fiber-return signal blocked. Shadow gates were extracted from BOTDR traces beyond the fiber-return window and receiver-corrected to estimate the residual spectral background. The averaged shadow-gate spectrum was corrected by $H_{\text{rx}}(f)$, providing the residual spectral background under low- and high-ER operation, as shown in Figure 1.b and Figure 1.c, respectively. This calibration step removes the receiver spectral response and allows the SMF and DCF Brillouin bands to be compared on the same corrected RF scale.

Figure *3* shows the calibrated RF spectral maps at ER = 18.5 and 33.0 dB. At low ER, the SMF band near 10.85 GHz extends into the DCF region, while the DCF band near 9.62 GHz is also observed outside its nominal section, indicating nonlocal scattering from the residual inter-pulse field. At high ER, these cross-section components are reduced and the DCF response becomes more spatially confined.

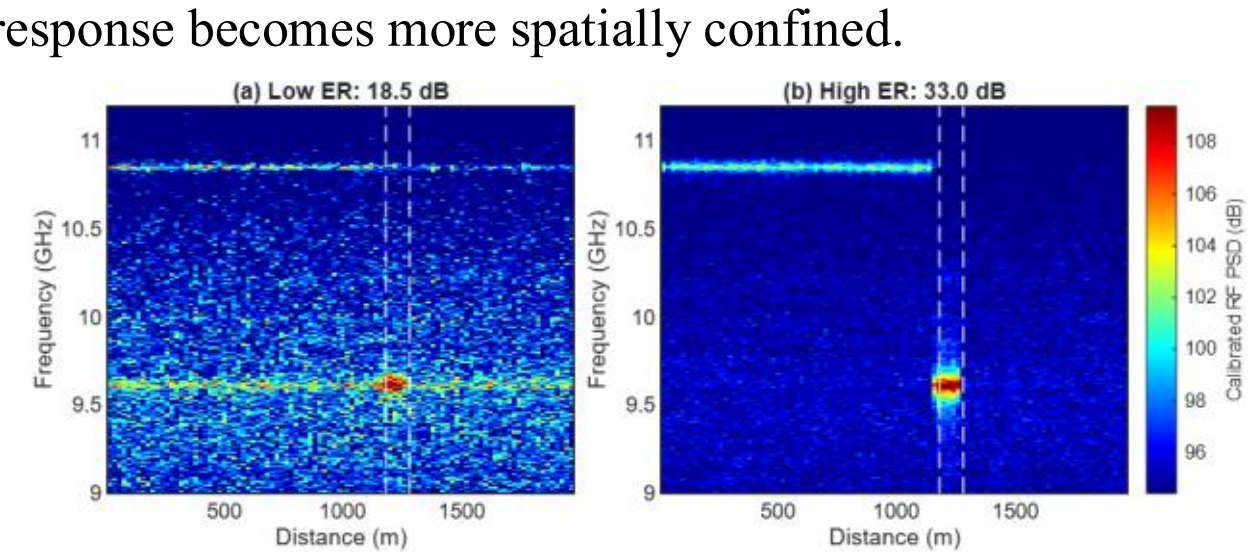

Figure 3. RF spectral power density versus distance and frequency for (a) ER = 18.5 dB, and (b) ER = 33.0 dB. Dashed vertical lines mark the SMF-DCF boundaries.

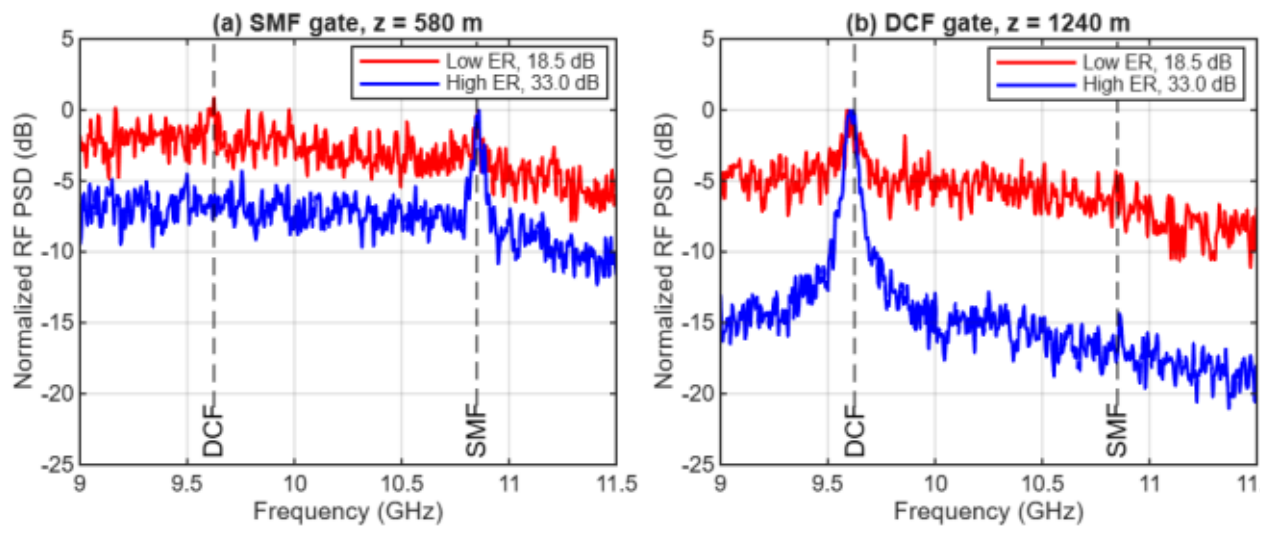

Figure 4. Representative RF spectra from (a) an SMF gate and (b) a DCF gate under low- and high-ER operation.

Figure *4* shows representative calibrated RF spectra from one SMF gate and one DCF gate at low and high ER. In the SMF gate, the local response is centered near 10.85 GHz and the DCF-band contribution near 9.62 GHz gives the contamination term, which is most evident at low ER. The opposite assignment applies in the DCF gate. These spectra define the bands used for SCR extraction.

For each spatial gate, the receiver-corrected spectrum was integrated over the SMF and DCF Brillouin bands at 10.85 and 9.62 GHz, using 30 and 45 MHz bandwidths, respectively. SCR was calculated as the contaminating-to-local band-power ratio. Figure *5* shows the resulting band powers along the link. At ER = 18.5 dB, the SMF-to-DCF contamination was $-5.3$ dB and decreased to $-14.8$ dB at ER = 33.0 dB, a reduction of 9.5 dB. Over the same ER change, the DCF-to-SMF contamination decreased from 2.9 to $-1.6$dB. The different residual levels reflect the unequal section lengths and Brillouin responses.

The extinction ratio was then varied by adjusting the MZM bias conditions, and the SCR was extracted across the SMF and DCF spatial gates. *Figure 6* shows the mean SCR values, with error bars indicating the standard deviation across the gates used for each fiber section. Over the measured ER range of 16-33 dB, the SCR decreased almost linearly with fitted slopes of -1.064 dB/dB for the SMF gates and -1.246 dB/dB for the DCF gates. These slopes are close to the -1 dB/dB dependence predicted by the finite-extinction leakage model, indicating that residual inter-pulse power is the main ER-dependent term under the fixed link and acquisition conditions. The fiber properties remain unchanged during the ER sweep and therefore principally determine the SCR offset and the difference between the two contamination directions. The fitted values are not universal coefficients: the unequal section lengths, integrated Brillouin responses, linewidths, attenuation, polarization-dependent reception, and integration bandwidths affect the absolute *SCR* levels. The larger deviation of the DCF slope is attributed to the smaller number of DCF gates, the different integration bandwidth, and the finite receiver background. These measurements validate the approximately inverse ER dependence for the experimentally tested condition of 200 ns pulses and approximately 20 m spatial resolution.

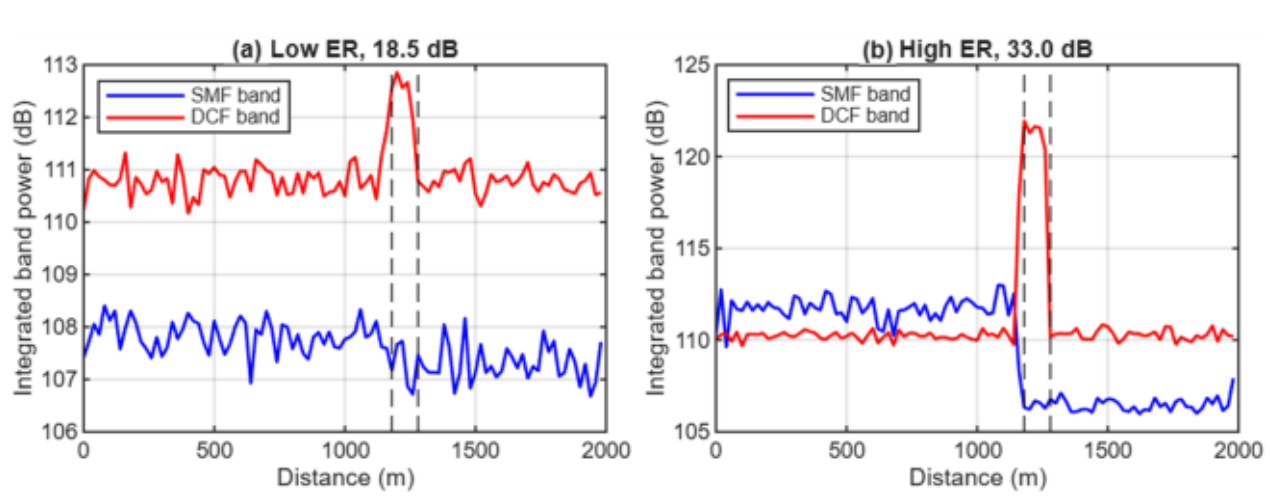

Figure 5. Integrated SMF- and DCF-band powers along the link for (a) low-ER and (b) high-ER operation.

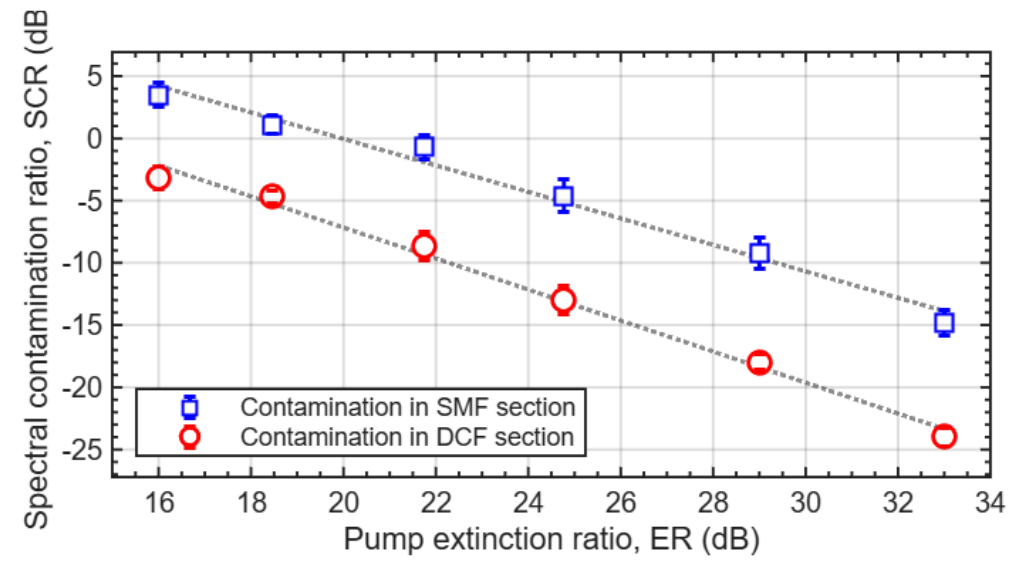

Figure 6. Mean SCR versus extinction ratio for SMF and DCF gate groups. Error bars show the standard deviation across spatial gates and the dashed lines are the linear fits.

## IV. Discussion

These measurements identify pulse extinction as a constraint on spatial locality in short-pulse BOTDR. Signal-processing approaches can improve SNR or acquisition speed but do not remove Brillouin leakage generated outside the interrogated gate [7]. A similar constraint occurs in phase-sensitive OTDR, where residual inter-pulse power generates Rayleigh noise and limits spatial resolution for a fixed sensing length [8]. Previous finite-ER analyses treated the residual inter-pulse Brillouin contribution mainly through its effect on SNR and the resulting BFS or temperature uncertainty [5]. In that framework, the leakage adds Brillouin power outside the intended pulse contribution and therefore increases the signal-dependent noise. SCR describes the same physical leakage from a different perspective. Instead of combining all leakage contributions into an overall SNR penalty, it quantifies the fraction of Brillouin power generated in one fiber section that is detected in gates assigned to another section. SNR and sensing uncertainty therefore describe the final measurement degradation, while SCR retains information on the origin and direction of the contamination. This distinction is particularly relevant near heterogeneous-fiber interfaces, where the total detected signal may remain usable even though part of the power assigned to a spatial gate is nonlocal. A direct conversion of SCR into BFS, temperature, or strain error remains system dependent because unresolved leakage, receiver noise, averaging, and the BFS estimator also contribute to the final uncertainty.

The SMF/DCF link is used here to separate section-to-section leakage through the BFS difference. Direct section-resolved SCR therefore requires sufficient spectral separation between the local and contaminating contributions. The different properties of the SMF and DCF affect the measured SCR levels, but remain unchanged throughout the ER sweep. Differences in Brillouin gain and effective mode area, represented through the effective Brillouin response, affect the Brillouin power generated per unit length and therefore the relative strength of the local and contaminating contributions. Section length and attenuation determine the effective length over which leakage accumulates and contribute to the asymmetry between the two contamination directions. Spectral linewidth also affects the integrated band power. In the present processing, the SMF and DCF bands are integrated over their measured linewidths of approximately 30 and 45 MHz, respectively. Dispersion is not an explicit first-order term in the integrated-power SCR model, although propagation-induced pulse-shape changes can contribute to the effective section response and remain fixed during the ER sweep. The BFS difference mainly provides the spectral separation needed to identify the section of origin. Since none of these fiber parameters changes with ER, they can explain the absolute SCR levels and the difference between the two contamination directions, but not the systematic SCR decrease observed during the ER sweep. The measured slopes should therefore be interpreted as the ER dependence for the present fixed SMF/DCF configuration, and not as universal fiber-independent values.

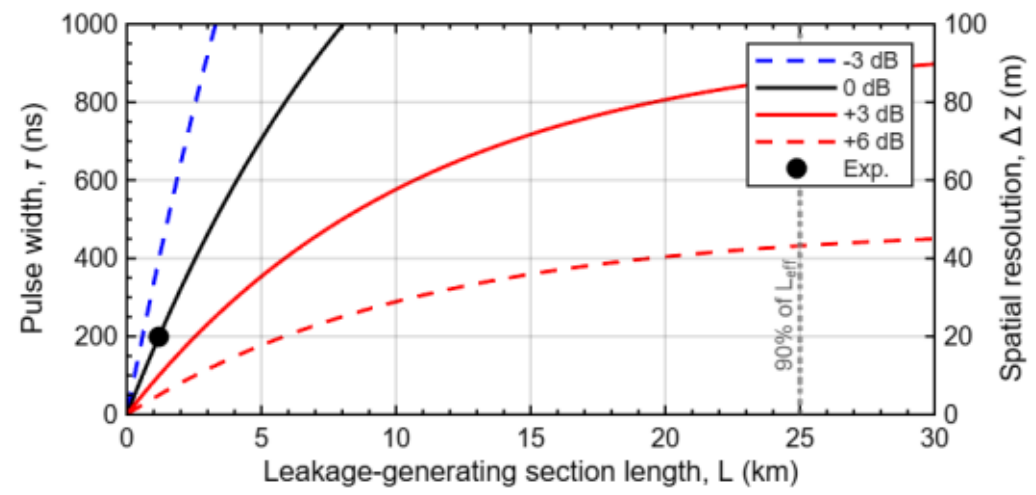

Figure 7. Calculated *ΔSCR*, referenced to the 1.2 km, 200 ns experimental condition.

In homogeneous SMF, same-BFS leakage can be estimated either from an ER sweep or from a high-ER reference trace. In the first case, ER is varied while the pulse width, on-state power, polarization, and receiver settings are kept fixed, and the ER-dependent excess Brillouin power is normalized to the local response. In the second case, a high-ER trace is taken as the local reference and the excess power measured at lower ER provides the contamination estimate. Both approaches quantify the contamination level, but not the exact position where the leakage was generated.

The dependence on pulse width and leakage-generating section length follows from the SCR model in (3)–(5). Since $\Delta z = c\tau/(2n)$, and $L_{\mathrm{eff}}(L) = [1 - \exp(-2\alpha L)]/(2\alpha)$ [5], the first-order dependence for fixed fiber and acquisition conditions becomes $SCR \propto L_{\mathrm{eff}}/(ER \cdot \tau)$. Taking $(L_{\mathrm{ref}}, \tau_{\mathrm{ref}})$ as a reference, the change in SCR at fixed ER can be calculated:

$$\Delta SCR_{\mathrm{dB}} = 10\log_{10}\left[\frac{L_{\mathrm{eff}}(L)}{L_{\mathrm{eff}}(L_{\mathrm{ref}})}\frac{\tau_{\mathrm{ref}}}{\tau}\right]. \qquad (6)$$

*Figure 7* shows this dependence using the experimental SMF condition of $L_{\mathrm{ref}} = 1.2\mathrm{km}$ and $\tau_{\mathrm{ref}} = 200\mathrm{ns}$, assuming 0.2 dB/km fiber attenuation, corresponding to $\alpha = 0.046\ \mathrm{km}^{-1}$in the effective-length expression. For the same leakage-generating length, reducing the pulse width from 200 to 100 ns increases the calculated SCR by 3.0 dB, while a reduction to 50 ns gives a 6.0 dB increase. These pulse widths correspond to spatial resolutions of 20, 10, and 5 m for $n = 1.5$. The effect of longer leakage-generating sections is also shown in *Figure 7*. At 25 km, a $+6$dB SCR change corresponds to a pulse width of about 430 ns, or 43 m spatial resolution. Increasing the pulse width to about 860 ns, or 86 m resolution, reduces this change to $+3$dB. The curves flatten at longer lengths because $L_{\mathrm{eff}}$ approaches its attenuation-limited value. At 25 km, $L_{\mathrm{eff}}$ has reached about 90% of this value, so further increases in section length have a smaller effect on SCR. The calculation shows the trade-off between spatial resolution, leakage-generating length, and finite-ER contamination. It assumes that the Brillouin response, polarization-dependent reception, integration bandwidth, and pulse shape remain unchanged, so *Figure 7* gives relative SCR changes for the specific fiber link.

The scaling in *Figure 7* provides a simple procedure for ER selection. First, an acceptable SCR limit is selected for the required sensing accuracy. Second, the ER required to reach this limit is obtained from the measured SCR–ER curve for the reference configuration. Third, the SCR increase associated with a different pulse width or leakage-generating section length is obtained from *Figure 7* and added as an ER margin. Since the measured SCR–ER slopes are close to $-1$dB/dB, this can be approximated as $ER_{\mathrm{min}} \approx ER_{\mathrm{ref}} + \Delta SCR_{\mathrm{dB}}$. As an example, for an SCR limit of $-15$dB, the fitted SCR–ER relation gives an estimated ER requirement of about 34 dB for the reference 200 ns case. The +3 and +6 dB penalties for 100 and 50 ns increase the corresponding estimated requirements to approximately 37 and 40 dB. These values are specific to the present link and selected SCR limit, while the same procedure can be applied to other configurations. At sufficiently high ER, the finite-leakage term becomes comparable to the residual background, and the SCR will approach a plateau set by receiver noise and EDFA ASE.

## V. Conclusion

We experimentally quantified finite-ER-induced spatial nonlocality in BOTDR using the spectral contamination ratio. Increasing ER from 18.5 to 33.0 dB reduced the SMF-to-DCF contamination from $-5.3$ to $-14.8$ dB, while SCR slopes of $-1.064$ and $-1.246$ dB/dB were measured over 16–33 dB. The results provide a basis for link-specific ER selection.